# Magnetic and Mössbauer studies of $Ca_x Zn_{1-x} Fe_2 O_4$ nanoferrites

Kemi Y. Adewale [1], Itegbeyogene P. Ezekiel [2]

**Abstract.** A one-step synthesis of $Ca_xZn_{1-x}Fe_2O_4$ ($x = 0$, $0.5$ and $1$) nanoferrites and nanocomposites by the glycol-thermal method is reported. The structural, morphological and magnetic properties were studied using XRD, HRTEM, HRSEM, Mössbauer spectroscopy and VSM. The XRD patterns show a single phase cubic spinel structure for *x*=0. A composite phase of a spinel and hematite-like structure was observed for *x*=0.5, and for *x*=1, CaFe$_2$O$_4$ has the same structure as hematite. The superparamagnetic nature of the samples was confirmed from the Mössbauer and magnetization results.

## 1. Introduction

The diverse synthesis techniques of ferrites have provided options to improve their physical properties and to expand their possible applications [1]. Ferrites belong to a special class of ceramic oxides with a general chemical formula of $MFe_2O_4$, where $M$ represents one or a mixture of the divalent transition metals such as Mn, Zn, Ni, Co, Cu, Fe, Mg or Ca [2]. A typical example is $Ca_xZn_{1-x}Fe_2O_4$ where, $0 \leq x \leq 1$. At both extreme conditions for $x$, $ZnFe_2O_4$ is obtained for $x = 0$ and CaFe$_2$O$_4$ for $x = 1$. An intermediate case is obtained for $x = 0.5$. Zinc ferrite ZnFe$_2$O$_4$ is well known to be a normal cubic spinel [3] while there are several reports on the orthorhombic structure of calcium ferrite CaFe$_2$O$_4$ [4]. Consequently, one could vary the structural and magnetic properties of the ferrite by establishing different intermediate cases to produce new composite.

This work reports on the study of the influence of Ca$^{2+}$ ions on the cubic spinel structure of ZnFe$_2$O$_4$. The biocompatibility and environmental friendliness of Zn and Ca informed their choice in this study. The nanocomposite of the spinel/hematite phases are also investigated.

## 2. Experimental details

Nanoferrites and nanocomposites of $Ca_xZn_{1-x}Fe_2O_4$ ($x = 0, 0.5$ and $1$) were prepared by the glycol-thermal method. High purity metal chlorides were used as the starting materials. The stoichiometric were dissolved in about 50 ml of de-ionized water and magnetically stirred. Ammonia solution (NH$_3$) was slowly added after about 10 minutes of stirring to initiate precipitation until the pH was about 10 and the solution was further stirred for 40 minutes. The precipitate was mixed with 100 ml of ethylene glycol and transferred into the reaction chamber of a Watlow PARR 4843 pressure reactor. The reactor was operated at a soak temperature of 200 °C for 6 hours, a pressure of 170 psi and angular speed of 300 rpm. The reacted product was washed several times with de-ionized water over a Whatman filter in a Büchner flask to remove the chlorides. The absence of chlorides was confirmed by the addition of standard solution of AgNO$_3$ to the filtrate. The samples were dried on the filter paper for 24 hours under a 200 W infrared light. The characterization of the samples was done at room temperature by XRD using a Co-Kα radiation, HRTEM, HRSEM, $^{57}$Fe Mössbauer spectroscopy and VSM.

## 3. Results and discussions

The refined XRD patterns for $Ca_xZn_{1-x}Fe_2O_4$ ($x = 0, 0.5$ and $1$) nanoparticles are shown in Figure 1. A pure cubic spinel phase of ZnFe$_2$O$_4$ was obtained for $x = 0$ [5]. Nanocomposite of a spinel and hematite-like structure of $Zn_{0.5}Fe_2O_4 / Ca_{0.5}Fe_2O_4$ is formed at $x = 0.5$ and a trigonal structure of $CaFe_2O_4$ is formed at $x = 1$. The formation of hematite-like structure is likely due to the large ionic radius of Ca$^{2+}$ [6] which is unable to be accommodated in the spinel structure and the preparation method. It is seen that such oxide phase does not exist for ZnFe$_2$O$_4$ where the ionic radius of Zn$^{2+}$ is much smaller than Ca$^{2+}$ [7]. The presence of Ca$^{2+}$ also increased the particle size as shown in the narrowing of XRD peaks for $x = 0.5$ and $1$ [8]. The crystallite size ($D$) was calculated from the most intense 311 peak for the spinel cubic phase and the 104 peak of the trigonal phase using the Debye–Scherrer formula, $D = K\lambda / W_{(hkl)} cos\theta$, where $K$ is a constant associated with the shape factor of 0.94 for spherical shapes, λ is the XRD wavelength, $W$ is the full width at half maximum, $\theta$ is the Bragg's angle and $hkl$ are the Miller indices of the selected peak [9]. The calculated values of the crystallite size $D$, the lattice parameters $a$ and the volume $V$ obtained from the refinements are presented in Table 1. Figure 2 and 3 shows the HRTEM and HRSEM morphologies. The images show that the particles are almost spherical in shape. The particle size increased significantly at $x = 0.5$ and $1$. The crystallinity of the samples is further confirmed from the lattice fringes shown in the inset of Figure 2. The magnified view of Figure 2b and Figure 2c shows that the particles of the former are bigger in size. This agrees with the results obtained from the XRD measurements. The room temperature $^{57}$Fe Mössbauer spectroscopy is presented in Figure 4. The data for $x = 0$ was best fitted with two sextets and two doublets. Three sextets and one doublet were used to fit $x = 0.5$ and 1. Three sextets and one doublet were used to fit $x = 0.5$ and 1. The two sextets for the sample $x = 0$ is associated with the coordination of $Fe^{3+}$ ions at tetrahedral A and octahedral B sites of the spinel crystal structure [10]. As seen in Table 2, the higher field of 470 kOe is attributed to the B sites and the lower field of 408 kOe is assigned to the A site [11]. The doublets account for the superparamagnetic state of the material. The third sextets in $x=0.5$ and 1 is assigned to the interstitial Fe coordination, which is neither sitting in A or B site. The large hyperfine magnetic field of 526 kOe and 520 kOe is attributed to the formation of hematite-like structure.

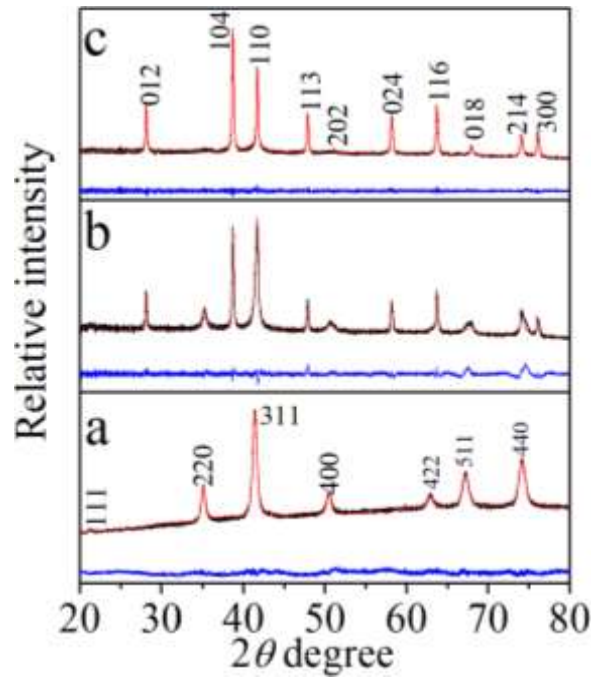

**Figure 1.** XRD patterns of $Ca_xZn_{1-x}Fe_2O_4$, (a) $x = 0$, (b) $x = 0.5$ and (c) $x = 1$

**Table 1.** Crystallite size, $D$, lattice parameters $a$ and volume $V$ for $Ca_xZn_{1-x}Fe_2O_4$ ($x = 0, 0.5$ and $1$).

| Sample | $D$ (nm) ±2 | $a$ (Å) ±0.0005 | $V$ (Å³) |
|---|---|---|---|
| ZnFe₂O₄ | 18 | 8.3981 | 592±0.06 |
| Zn₀.₅Fe₂O₄/Ca₀.₅Fe₂O₄ | 19 | $a = b = c = 8.3763$ | 588±0.06 |
|  | 50 | $a = b = 5.0295$ |  |
| | | $c$ = 13.7702 | 302±0.04 |
| CaFe$_2$O$_4$ | 45 | $a = b$ = 5.02962 | 302±0.04 |
| | | $c$ = 13.7743 | |

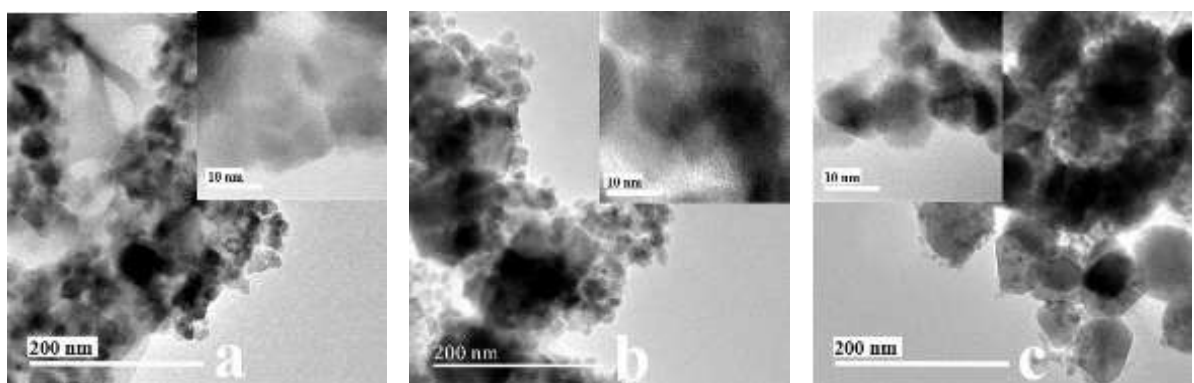

**Figure 2.** HRTEM images of $Ca_xZn_{1-x}Fe_2O_4$, (a) $x=0$, (b) $x=0.5$ and (c) $x=1$.

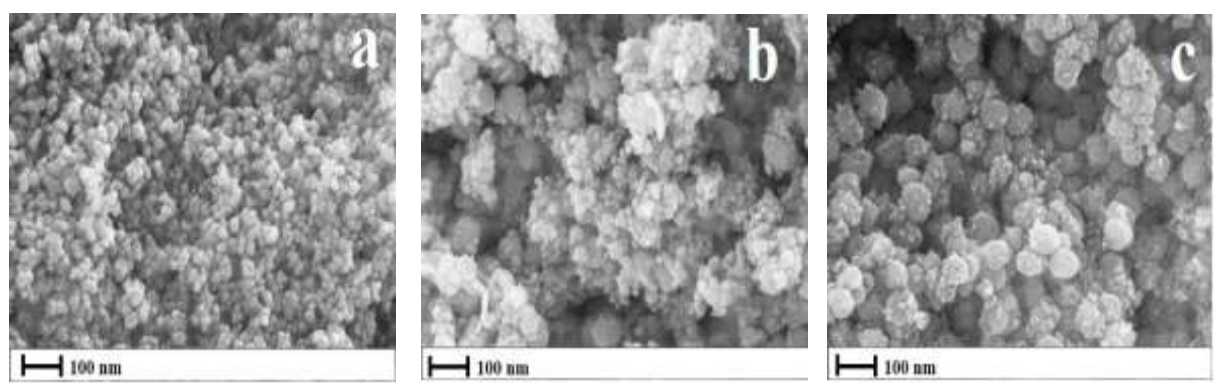

**Figure 3.** HRSEM images of $Ca_xZn_{1-x}Fe_2O_4$, (a) $x=0$, (b) $x=0.5$ and (c) $x=1$.

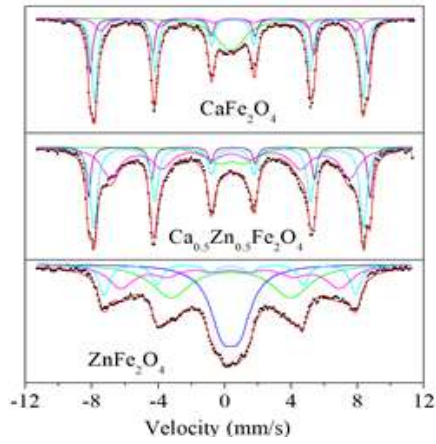
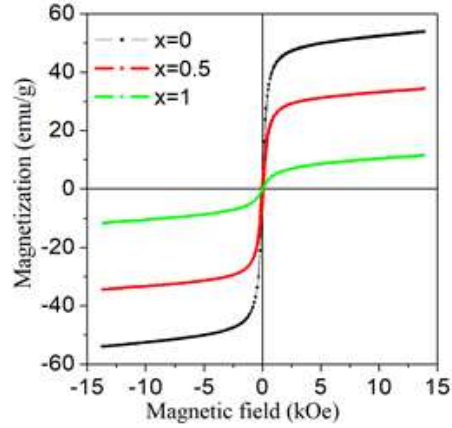

**Figure 4**. Fitted room temperature $^{57}$Fe Mössbauer spectra of $Ca_xZn_{1-x}Fe_2O_4$ ($x = 0, 0.5$ and $1$) nanoferrites

**Figure 5**. Room temperature hysteresis loops of $Ca_xZn_{1-x}Fe_2O_4$ ($x = 0, 0.5$ and $1$) nanoferrites

**Table 2.** Isomer shift ($\delta$), quadrupole splitting ($\Delta_{EQ}$), line widths ($LW$), fraction populations ($f$) and magnetic hyperfine field ($B_{hf}$) of the samples $Ca_xZn_{1-x}Fe_2O_4$ ($x = 0, 0.5$ and $1$).

| $x$ | Pattern | $\delta$ (mm/s) | | $\Delta_{EQ}$ (mm/s) | | $LW$ (mm/s) | | $f$ (%) | | $B_{hf}$ (kOe) |
|---|---|---|---|---|---|---|---|---|---|---|
| | | ±0.004 | ±0.002 | ±0.02 | ±0.001 | ±0.15 | ±0.004 | ±0.02 | ±0.03 | ±0.13 |
| 0 | B | 0.37 | 0.31 | 7.22 | -0.002 | 1.4 | 0.37 | 30 | 15 | 470 |
| | A | 0.35 | 0.32 | 1.03 | 0.000 | 0.88 | 0.95 | 32 | 24 | 408 |
| 0.5 | C$_1$ | 0.41 | 0.39 | 1.56 | -0.11 | 0.85 | 0.16 | 11 | 15 | 526 |
| | B | | 0.35 | - | -0.02 | - | 0.68 | - | 37 | 448 |
| | A | | 0.36 | - | -0.11 | - | 0.25 | - | 37 | 503 |
| 1 | C$_1$ | 0.36 | 0.39 | 0.42 | -0.11 | 0.98 | 0.18 | 21 | 25 | 520 |
| | B | - | 0.36 | - | -0.11 | - | 0.21 | - | 44 | 501 |
| | A | - | 0.38 | - | -0.16 | - | 0.35 | - | 11 | 477 |

**Table 3**. Coercivity $H_C$, saturation magnetization $M_S$, remanence magnetization $M_R$ and squareness ratio ($M_R/M_S$) obtained at room temperatures in applied field of 14 kOe for $Ca_xZn_{1-x}Fe_2O_4$ ($x = 0, 0.5$ and $1$)

| $x$ | $H_C$ (Oe) | $M_S$ (emu/g) | $M_R$ (emu/g) | $M_R/M_S$ |
|---|---|---|---|---|
| | ±0.55 | ±0.09 | ±0.44 | ±0.01 |
| 0 | 26 | 49.58 | 2.96 | 0.06 |
| 0.5 | 51 | 31.15 | 3.17 | 0.10 |
| 1 | 15 | 8.61 | 0.09 | 0.02 |

Figure 5 shows the room temperature magnetization measurements performed in a magnetic field of up to 14 kOe. In Table 3 we presented the coercive fields ($H_C$), saturation magnetizations ($M_S$), remanent magnetizations ($M_R$) and squareness of the loops ($M_R/M_S$) calculated from the hysteresis loops. The $M_S$ were estimated from the law of approach to saturation magnetization using

$$M(H) = M_S\left(1 - \frac{a}{H} - \frac{b}{H^2}\right) + \chi H,$$

where $M_S$, $a$, $b$ and $\chi$ are the best fit parameters to the data [12].

The superparamagnetic behavior of the samples is reflected in the low squareness ($M_R/M_S$ ratios) and in the small $H_C$ values [13]. It is observed that the saturation magnetization decreases from 50 emu/g at $x=0$ to 31 emu/g and 9 emu/g for $x=0.5$ and 1 respectively. The decrease in $M_S$ is due to the formation of hematite-like phase [14].

## 4. Conclusions

We synthesized nanoferrites and nanocomposites of $Ca_xZn_{1-x}Fe_2O_4$ ($x = 0, 0.5$ and $1$) at the room temperature via the glycol thermal method. The refined XRD patterns confirmed a single phase cubic spinel structure for $x=0$, a composite of a cubic/trigonal structure for $x=0.5$ and 1. The HRTEM and HRSEM image show clear differences in the morphologies of the samples. The substitution of $Zn^{2+}$ for $Ca^{2+}$ increased the crystallite size of the sample. [57]Fe Mössbauer spectroscopy measurements reveal the formation of hematite-like structure for $x=0.5$ and 1. This caused a decreased in the saturation magnetization for $x=0.5$ and 1.

## 5. Acknowledgements

The authors express gratitude to the National Research Foundation and the University of KwaZulu-Natal, South Africa for research grants.

[1,2] School of Chemistry and Physics, Westville Campus, University of KwaZulu-Natal, Private Bag X54001, Durban, South Africa

Corresponding Author:

Email: kemiadewale2@gmail.com (Kemi Y. Adewale)